\documentclass[12pt]{iopart}

\usepackage{amsfonts}

\usepackage[dvips]{graphicx}
\usepackage{iopams}
\usepackage{amsfonts}
\usepackage{cite}
\newcommand{\me}{\mathrm{e}}

\begin{document}

\title{Topological Quantum Gate Construction by Iterative Pseudogroup
  Hashing}

\author{Michele Burrello}
\address{International School for Advanced Studies (SISSA),
  Via Bonomea 265, 34136 Trieste, Italy}
\address{Istituto Nazionale di Fisica Nucleare, Sezione di Trieste, Italy}

\author{Giuseppe Mussardo}
\address{International School for Advanced Studies (SISSA),
  Via Bonomea 265, 34136 Trieste, Italy}
\address{Istituto Nazionale di Fisica Nucleare, Sezione di Trieste, Italy}
\address{International Centre for Theoretical Physics (ICTP),
  I-34014 Trieste, Italy}

\author{Xin Wan}
\address{Asia Pacific Center for Theoretical Physics, 
Pohang, Gyeongbuk 790-784, Korea}
\address{Department of Physics, Pohang University of Science and
Technology, Pohang, Gyeongbuk 790-784, Korea}
\address{Zhejiang Institute of Modern Physics, Zhejiang
University, Hangzhou 310027, P.R. China}

\date{\today}

\begin{abstract}

  We describe the hashing technique to obtain a fast approximation of
  a target quantum gate in the unitary group SU(2) represented by a
  product of the elements of a universal basis. The hashing exploits
  the structure of the icosahedral group [or other finite subgroups of
  SU(2)] and its pseudogroup approximations to reduce the search
  within a small number of elements. One of the main advantages of the
  pseudogroup hashing is the possibility to iterate to obtain more
  accurate representations of the targets in the spirit of the
  renormalization group approach. We describe the iterative
  pseudogroup hashing algorithm using the universal basis given by the
  braidings of Fibonacci anyons. The analysis of the efficiency of the
  iterations based on the random matrix theory indicates that the
  runtime and the braid length scale poly-logarithmically with the
  final error, comparing favorably to the Solovay-Kitaev algorithm.

\end{abstract}

\maketitle

\section{Introduction}

The possibility of physically implementing quantum computation opens
new scenarios in the future technological development. This justifies
the deep efforts made by the scientific community to develop
theoretical approaches suitable to effectively build a quantum
computer and, at the same time, to reduce all the sources of errors
that would spoil the achievement of quantum computation.

To implement quantum computation one needs to realize, through
physical operations over the qubits, arbitrary unitary operators in
the Hilbert space that describes the system. This task can be achieved
by using a finite number of elementary gates that constitutes a
basis. A small set of gates is said to be {\em universal} for quantum
computation if it allows to approximate, at any given accuracy, every
unitary operator in terms of a quantum circuit made of only those
gates~\cite{nielsen}. It can be shown that a basis able to reproduce
every SU(2) operator and one entangling gate (as CNOT) for every
pair of qubits is indeed universal~\cite{divincenzo}; therefore the
problem of finding an approximation of unitary operators in $SU(N)$
can be reduced in searching an efficient representation, in terms of
the elements of the basis, of single-qubit gates in SU(2) and of the
two-qubit gate CNOT.

For SU(2) it is possible to find a universal set of single-qubit
operators involving just two elementary gates, which we will label as
$\sigma_1$ and $\sigma_2$ (not to be confused with Pauli matrices!),
and their inverses, $\sigma_1^{-1}$ and $\sigma_2^{-2}$. This means
that every single-qubit gate can be efficiently approximated by a
string of these four elementary elements. In the scheme of topological
quantum computation, these fundamental gates are realized by
elementary braid operation on the excitations of the system, the
anyons. In order to be universal, the group obtained by multiplying
the four $\sigma$ gates must be dense in SU(2): it is therefore
sufficient that $\sigma_1$ and $\sigma_2$ do not belong to the same
finite subgroup of SU(2). For what concerns controlled gates in
$SU(4)$ as CNOT, their approximation can be usually reduced to the
case of operators in SU(2), as described in Refs.~\cite{hormozi07}
and \cite{xu082} in the context of topological quantum computation,
the main subject of our following considerations.

The simplest way to obtain an approximation of a given target gate $T
\in SU(2)$ using only four elementary gates $\sigma_{1,2}^{\pm 1}$ is
to search, among all the ordered products of $N$ of such operators,
the one which best represents $T$ minimizing its distance to it (the
rigorous definition of distance is given in section \ref{SU2}). This
operation is called the {\em brute-force} procedure \cite{hormozi07}.
The number of all the possible products of this kind grows
exponentially in $N$ as $\alpha^{N}$ (where $\alpha \approx 3$) and, because
of the three-dimensional nature of SU(2), one can show that, for a
suitable choice of the universal basis, the average error obtained
with different targets decreases approximately as $e^{-\frac{N}{3} \ln
 \alpha }$ (in section \ref{BFfibo} we will describe in more detail the
brute-force search for Fibonacci anyons).

This approach, consisting of a search algorithm over all the possible
ordered product up to a certain length, has an extremely clear
representation if one encodes qubits using non-abelian anyons. In this
case, the computational basis for the quantum gate is the set of the
elementary braidings between every pair of anyons, and their products
are represented by the braids describing the world lines of these
anyonic quasiparticles. Starting from this kind of universal basis,
the search among all the possible products of $N$ elements gives, of
course, the optimal result, but the search time is exponential in $N$
and therefore it is impractical to reach sufficiently small error for
arbitrary gates.

There are, however, other possible approaches that allow to reach an
arbitrary small error in a faster way, even if they do not obtain the
best possible result in terms of the number $N$ of elementary gates
involved. The textbook example is the Solovay-Kitaev algorithm
\cite{nielsen, kitaev, dawson06}. This algorithm provides a powerful
tool to obtain an approximation of arbitrary target gates in SU(2) at any
given accuracy starting from an $\epsilon$-net, i.e. a finite covering
of SU(2) such that for every single-qubit operator $T$ there is at
least one gate inside the $\epsilon$-net that has a distance from $T$
smaller than $\epsilon$. The Solovay-Kitaev algorithm is based on
the decomposition of small rotations with elements of the
$\epsilon$-net and both the runtime and the length of the final
product of elementary gates scale poly-logarithmically with the final
error $\varepsilon$. The exponents depend on the detailed construction
of the algorithm: the simplest realization of the algorithm, as
provided in \cite{hormozi07} in the context of topological quantum
computation, is characterized by the following scaling:
\begin{eqnarray} \label{NKS}
N\sim\left( \ln \left(1/\varepsilon \right) \right)^c \qquad {\rm with} 
\quad c= \frac{\ln 5}{\ln \left(3/2 \right) }\approx 3.97 \\
T \sim \left( \ln \left(1/\varepsilon \right) \right)^d 
\qquad {\rm with} \quad 
d = \frac{\ln 3}{\ln \left(3/2 \right) }\approx 2.71
\end{eqnarray}
As discussed in the Appendix 3 of \cite{nielsen} and in
\cite{dawson06}, a more sophisticated implementation of the
Solovay-Kitaev algorithm realizes
\begin{eqnarray}
N \sim \left( \ln \left(1/\varepsilon \right)\right)^2 
\ln\left(\ln \left(1/\varepsilon \right)\right) \label{LKS} \\
T \sim \left( \ln \left(1/\varepsilon \right)\right)^2 
\ln\left(\ln \left(1/\varepsilon \right)\right). \label{TKS}
\end{eqnarray}

The hashing algorithm, which was proposed in Ref.~\cite{burrello}, has
the aim of obtaining a more efficient approximation of a target
operator than the Solovay-Kitaev algorithm in a practical regime, with
a better time scaling and without the necessity of building an
$\epsilon$-net covering the whole target space of unitary
operators. Our main strategy will be to create a dense mesh
$\mathcal{S}$ of fine rotations with a certain average distance from
the identity and to reduce the search of the target approximation to
the search among this finite set of operators that, in a certain way,
play the role of the $\epsilon$-net in a neighborhood of the
identity. This set will be built exploiting the composition property
of a finite subgroup (the icosahedral group) of the target space
SU(2) and exploiting also the almost random errors generated by a
brute-force approach to approximate at a given precision the elements
of this subgroup. Therefore the algorithm allows us to associate a
finite set of approximations to the target gate and each of them is
constituted by an ordered product of the elementary gates chosen in a
universal quantum computation basis. Since our braid lookup task is
similar to finding items in a database given its search key, we borrow
the computer science terminology to name the procedure {\em hashing}.

With this algorithm we successfully limit our search within a small
number of elements instead of an exponentially growing one as in the
case of the brute-force search; this significantly reduces the runtime
of the hashing algorithm. Furthermore, we can easily iterate the
procedure in the same spirit of the renormalization group scheme,
based on the possibility of restricting the set $S$ in a smaller
neighborhood of the identity at each iteration, obtaining in this way
a correction to the previous result through a denser mesh.

In Sec.~\ref{topo} we briefly review the main ideas of topological
quantum computation, which is the natural playground for the
implementation of the hashing procedure. we use Fibonacci anyons to
encode qubits and define their braidings operators that are the main
example of universal elementary gates that we use to analyze our
algorithm. In Sec.~\ref{SU2} we introduce the basic ideas of the
icosahedral group and its braid representations, which are
pseudogroups. We describe in detail the iterative hashing algorithm in
Sec.~\ref{hashing} and analyze performance of its iteration scheme in
Sec.~ref{efficiency}. We conclude in Sec.~\ref{sec:conclusion}. In
\ref{sec:weave} we derive the distribution of the best approximation
in a given set of braids, which can be used to estimate the
performance of, e.g., the brute-force search.

\section{Topological quantum computation and Fibonacci
  anyons} \label{topo}

The discovery of condensed matter systems which show topological
properties and cannot be described simply by local observables has
opened a new perspective in the field of quantum
computation. Topological quantum computation is based on the
possibility of encoding the qubits necessary for quantum computation
in topological properties of physical systems, and obtaining in such a
way a fault-tolerant computational scheme protected by local noise
\cite{kitaev03,freedman02,nayak08,brennen08,preskill}.

For topological quantum computation one needs anyons with non-abelian
statistics. Unlike fermions or bosons, anyons are quasiparticles whose
exchange statistics is described not by the permutation group, but by
the braid group, generated by the elementary exchanges of anyon
pairs. In particular, in certain two-dimensional topological states of
matter, a collection of non-Abelian anyonic excitations with fixed
positions spans a multi-dimensional Hilbert space and, in such a
space, the quantum evolution of the multi-component wavefunction of
the anyons is realized by braiding them.

Therefore, it is natural to consider the unitary matrices representing
the exchange of two anyons as the elementary gates for a quantum
computation scheme. In this way the universal basis for the quantum
computation acquires an immediate physical meaning and its elements
are implemented in a fault-tolerant way; therefore the problem of
approximating a target unitary gate is translated in finding the best
``braid'' of anyons that represents the given operator up to a certain
length of ordered anyons exchanges \cite{hormozi07}.

There are several physical systems characterized by a topological
order that are suitable to present non-abelian anyonic
excitations. The main experimental candidates are quantum Hall systems
in semiconductor devices (see \cite{stern} for an introduction to the
subject), as well as similar strongly correlated states in cold atomic
systems \cite{cooper01}, $p_x+ip_y$ superconductors \cite{read00}, and
superconductor-topological insulator heterostructures \cite{nayak10}.

One of the most studied anyonic models is the Fibonacci anyons model
(see for example \cite{bonesteel05,simon06,hormozi07,xu08,baraban10})
so named because the dimension of their Hilbert space follows the
famous Fibonacci sequence, which imply that their quantum dimension is
the golden ratio $\varphi = (\sqrt{5} + 1)/2$. Fibonacci anyons
(denoted by $\phi$ as opposed to the identity $1$) are known to have a
simple fusion algebra ($\phi \times \phi = \phi + 1$) and to support
universal topological quantum computation~\cite{freedman02}. Candidate
systems supporting the Fibonacci fusion algebra and braiding matrices
include fractional quantum Hall states known as the Read-Rezayi state
at filling fraction $\nu = 3/5$~\cite{read99} (whose particle-hole
conjugate is a candidate for the observed $\nu = 12/5$ quantum Hall
plateau~\cite{xia04}) and the non-Abelian spin-singlet states at $\nu
= 4/7$~\cite{ardonne99}.

Every anyonic model, as the one of Fibonacci anyons, is characterized
by several main components: the superselection sectors of the theory
are the different kinds of anyons which are present ($1$ and $\phi$ in
our case), their behaviour is described by the fusion and braiding
rules that are linked through the associativity rules of the related
fusion algebra, expressed by the so called $F$ matrices (see Refs.
\cite{preskill,kitaev06} for a general introduction to the anyonic
theories and Ref.~\cite{hormozi07} for its application to the
Fibonacci case).

If we create two pairs of $\phi$s out of the vacuum, both pairs must
have the same fusion outcome, $1$ or $\phi$, forming a qubit;
therefore the logical value of the qubit is associated to the result
of the fusion of the two Fibonacci anyons inside the first or the
second pair, while the total fusion outcome is the vacuum. The
braiding of the four $\phi$s can be generated by two fundamental
braiding matrices (related by the Yang-Baxter equation)

\begin{equation}
\label{eq:s2}
\sigma_1 = \left [ 
\begin{array}{cc}
\me^{-i 4 \pi / 5} & 0 \\
0 & -\me^{-i 2 \pi / 5}
\end{array}
\right ], 
\end{equation}
\begin{equation}
\sigma_2 = F^{-1} \sigma_1 F= \left [ 
\begin{array}{cc}
-\tau \me^{-i \pi / 5} & -\sqrt{\tau} \me^{i 2 \pi / 5} \\
 -\sqrt{\tau} \me^{i 2 \pi / 5} & -\tau
\end{array}
\right ], 
\end{equation}

\noindent
and their inverses $\sigma_1^{-1}$, $\sigma_2^{-1}$. Here $\tau =
\varphi^{-1}=(\sqrt{5} - 1)/2$. We notice that $\sigma_1$ is the
elementary braiding involving two anyons in the same pair, thus it is
diagonal in the qubit basis; instead, $\sigma_2$ describes the
braiding of anyons in different pairs and can be obtained by the
change of basis defined by the associativity matrix $F$.

This matrix representation of the braidings generates a four-strand
braid group $B_4$ (or an equivalent three-strand braid group $B_3$):
this is an infinite dimensional group consisting of all possible
sequences of length $L$ of the above generators and, with increasing
$L$, the whole set of braidings generates a dense cover of the SU(2)
single-qubit rotations.

Besides, Simon {\it et al.} \cite{simon06} demonstrated that, in order
to achieve universal quantum computation, it is sufficient to move a
single Fibonacci anyon around the others at fixed position. Thanks to
this result, one can study an infinite subgroup of the braid group $B$
which is the group of \textit{weaves}, braids characterized by the
movement of only one quasiparticle around the others. From a practical
point of view it seems simpler to realize and control a system of this
kind, therefore we will consider only weaves in the following. There
is also another advantage in doing so: the elementary gates to cover
SU(2) become $\sigma_1^2$, $\sigma_2^2$ and their inverses;
therefore the weaves avoid the equivalence between two braids caused
by the Yang-Baxter relations, so that it is more immediate to find the
set of independent weaves up to a certain length. In fact, the only
relations remaining for Fibonacci anyons that give rise to different
but equivalent weaves are the relations
$\sigma_1^{10}=\sigma_2^{10}=1$, because they imply that $\sigma_i^6 =
\sigma_i^{-4}$ and one can always reduce weaves with terms of the kind
$\sigma_i^{\pm 6}$ to shorter but equivalent ones.

In order to calculate the efficiency in approximating a target gate
through the brute-force search, which gives the optimal result up to a
certain length, it will be useful calculating here the number of
independent weaves of Fibonacci anyons; in general a weave of length
$L$ can be written as
\begin{equation}
\sigma_{p_1}^{q_1} \sigma_{p_2}^{q_2} \cdots \sigma_{p_s}^{q_s}, 
\end{equation}
where $\ p_i \in \{1, 2\}$, $p_i \neq p_{i \pm 1}$, and $\sum_{i}
\vert q_i \vert = L$. As mentioned above, to ensure that this braid is
not equivalent to a shorter braid, one must have $q_i = \pm 2$ or $\pm
4$.

Let us assume the number of length-$L$ weaves is $N(L)$, consisting of
$N_4(L)$ weaves ending with $\sigma_p^{\pm 4}$ and $N_2(L)$ weaves
ending with $\sigma_p^{\pm 2}$. To form a weave of length $L + 2$, the
sequence must be appended by $\sigma_r^2$ or $\sigma_r^{-2}$. For
sequences ending with $\sigma_p^2$ (or $\sigma_p^{-2}$), we can append
$\sigma_p^2$ (or $\sigma_p^{-2}$), $\sigma_{3-p}^2$, or
$\sigma_{3-p}^{-2}$. However, for sequences ending with $\sigma_p^4$
(or $\sigma_p^{-4}$), we can only append $\sigma_{3-p}^2$ or
$\sigma_{3-p}^{-2}$. Therefore, we have the recurrence relations:
\begin{eqnarray}
N(L) = N_4(L) + N_2(L), \\
N(L + 2) = 2 N_4(L) + 3 N_2(L), \\
N_4(L + 2) = N_2(L).
\end{eqnarray}
With an ansatz $N(L) \sim \alpha^{L/2}$ [and so are $N_4(L)$ and
  $N_2(L)$], we find $\alpha = 1 \pm \sqrt{3}$. Therefore the exact
number of weaves of length $L$ is
\begin{equation} \label{weaves}
N(L) = \left(1-\frac{1}{\sqrt{3}}\right)
   \left(1-\sqrt{3}\right)^{L/2}+\left(1+\frac{1}{\sqrt{3}}
\right) \left(1+\sqrt{3}\right)^{L/2},
\end{equation}
which grows as $(1 + \sqrt{3})^{L/2}$ asymptotically.

\section{SU(2), subgroups and pseudogroups} \label{SU2}

\subsection{Single-qubit gates and distances in SU(2)}
The hashing algorithm allows to approximate every target single-qubit
gate in SU(2) exploiting the composition rules of one of its
subgroups, as the icosahedral one; therefore it is useful to consider
the standard homomorphism from the group of rotations in
$\mathbb{R}^3$, $SO(3)$, and the group of single-qubit gates, that
permits to write every operator $U \in SU(2)$ as
\begin{eqnarray} 
\fl
U\left(\hat{m},\phi \right)  = e^{i \hat{m} \cdot \vec{\sigma} (\phi/2)} = \nonumber \\
= \left (
\begin{array}{cc}
\cos (\phi/2) + i m_z \sin(\phi/2) & m_y \sin(\phi/2) + i m_x
\sin(\phi/2) \\ - m_y \sin(\phi/2) + i m_x \sin(\phi/2) & \cos
(\phi/2) - i m_z \sin(\phi/2)
\end{array}
\right ).
\label{oper}
\end{eqnarray}
$U\left(\hat{m},\phi \right)$ represents a rotation, in $SO(3)$, of an
angle $\phi$ around the axes identified by the unitary vector
$\hat{m}$. Therefore, if we exclude an overall phase, which is
unimportant to the purpose of single-qubit gates, SU(2) can be
mapped on a sphere of radius $\pi$: a point in the sphere defined by a
radius $0 \le \phi \le \pi$ in the direction $\hat{m}$ corresponds to
the rotation $U(\hat{m},\phi)$. In the following we will often address the elements of SU(2) not only as single qubit gates but also as
rotations, implicitly referring to this homomorphism.

The distance $d$ (also referred to as error) between two gates (or
their matrix representations) $U$ and $V$ is defined as the operator
norm distance
\begin{equation}
d\left( U,V\right) \equiv \Vert U - V \Vert = \sup _{\Vert \psi \Vert = 1} \Vert \left( U-V\right) \psi
  \Vert.
\end{equation}
Thus, if we consider two operators $U=U(\hat{m},\phi)$ and
$V=U(\hat{n},\theta)$ the distance between them is
\begin{equation} \label{dist}
 d\left( U,V\right)=\sqrt{2-2\cos{\frac{\phi}{2}}\cos{\frac{\theta}{2}} -2\,\hat{m}\cdot\hat{n}\,\sin{\frac{\phi}{2}}\sin{\frac{\theta}{2}}},
\end{equation}
which is bound above by $\sqrt{2}$.  We notice that the distance
of a rotation $U(\hat{m},\phi)$ from the identity operator is $d =
2\sin{({\phi}/{4})}$.

\subsection{Brute-force search with Fibonacci anyons} \label{BFfibo}

First, we estimate the efficiency for the brute-force search algorithm
through the probability distribution of the error of any weave of
length $L$ from a given target gate in SU(2). We assume that the
collection of nontrivial weaves of length $L$, whose number is $N(L)$
as given in Eq.~(\ref{weaves}), are uniformly distributed on the
sphere of SU(2). Representing the elements of SU(2) on the surface
of the four-dimensional sphere, we find the distribution $p_{BF}(d)$
of their distance from the identity operator to be
\begin{equation} \label{pBF}
p_{BF}(d)=\frac{4}{\pi}\,d^2\sqrt{1-\left( d/2\right)^2 }
\end{equation}
where $d = 2\sin{({\phi}/{4})}$.

Given this distribution, we obtain the average error for the
brute-force search to be 
\begin{equation} \label{BFerror}
\bar{d}(L) = \frac{\pi^{1/3} \Gamma
   \left(\frac{1}{3}\right)}{6^{2/3} [N(L)]^{1/3}}
\approx 1.021 e^{-L/5.970}
\end{equation}
asymptotically (see \ref{sec:weave} for more detail).

\begin{figure}[ht]
  \centering
  \includegraphics[width=10cm]{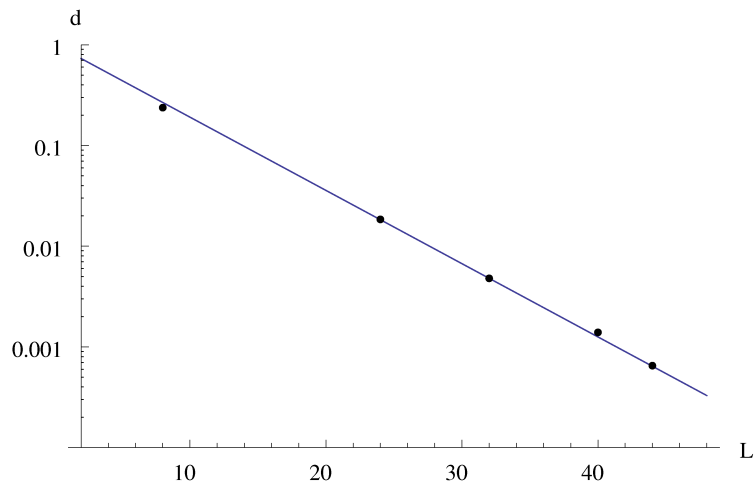}
  \caption{The average errors (dots) of the approximations to the 60
    rotations of the icosahedral group with the brute-force search as
    a function of the length of the weaves used. These errors
    characterize the pseudogroups used in the hashing algorithm for
    the lengths $L = 8$, 24, 32, 40, and 44. The results are in very
    good agreement with the prediction in Eq.~(\ref{BFerror}) (solid
    line).}
  \label{fig:bf}
\end{figure}

\subsection{The Icosahedral group}

The hashing procedure relies on the possibility of exploiting
the group structure of a finite subgroup of the target space to build
sets $\mathcal{S}(L)$ of fine rotations, distributed with smaller and
smaller mean distances from the identity, that can be used to
progressively correct a first approximation of a target gate
$T$. Therefore, it is fundamental to search, for every target space of
unitary operators, a suitable subgroup to build the sets
$\mathcal{S}$. Among the different finite subgroups of SU(2) we
considered the $SO(3)$ subgroups corresponding to the symmetry group
of the icosahedron and of the cube, which have order 60 and 24,
respectively. However, for practical purposes, we will refer in the
following mainly to the icosahedral group because its
implementation of the hashing algorithm, as we will describe
below, is more efficient in terms of the final braid length.

The icosahedral rotation group ${\mathcal I}$ is the largest
polyhedral subgroup of SU(2), excluding reflection. For this reason, it
has been often used to replace the full SU(2) group for practical
purposes, as for example in earlier Monte Carlo studies of SU(2)
lattice gauge theories~\cite{rebbi80}, and its structure can be
exploited to build meshes that cover the whole SU(2)
\cite{mosseri08}.

${\mathcal I}$ is composed by 60 rotations around the axes of symmetry
of the icosahedron (platonic solid with twenty triangular faces) or of
its dual polyhedron, the dodecahedron (regular solid with twelve
pentagonal faces); there are six axes of the fifth order
(corresponding to rotations with fixed vertices), ten of the third
(corresponding to the triangular faces) and fifteen of the second
(corresponding to the edges). Let us for convenience write ${\mathcal
  I} = \{g_0, g_1, ...,g_{59}\}$, where $g_0 = e$ is the identity
element.  In figure \ref{fig:pseudogr}\textbf{a} the elements of the
icosahedral group are represented inside the SU(2) sphere.

Because of the group structure, given any product of $n$ elements of a
subgroup $g_{i_1}g_{i_2}\cdots g_{i_n}$ there always exists a group
element $g_{i_{n+1}} = g_{i_n}^{-1}\cdots g_{i_2}^{-1}g_{i_1}^{-1}$
that is its inverse. In this way $g_{i_1}g_{i_2}\cdots
g_{i_n}g_{i_{n+1}}=e$ and one can find $O^n$ different ways of
obtaining the identity element, where $O$ is the order of the group
(60 in the case of the icosahedral one).  This is the key property we
will use in order to create a dense mesh $\mathcal{S}$ of fine
rotations distributed around the identity operator in the target
space. To achieve this goal, however, we need to break the
exact group structure and to exploit the errors given by a brute-force
search approximation of the elements of the chosen subgroup.

\subsection{Pseudogroup}

The main idea in the realization of the mesh $\mathcal{S}$ is that,
representing the 60 elements of the subgroup $\mathcal{I}$ with weaves
of a given length $L$, we can control, due to the relation
(\ref{BFerror}), the average distance (or error) between the exact
rotations in $\mathcal{I}$ and their braid representations that
constitute the set $\tilde{\mathcal{I}}(L)$ which we will refer to as
a \textit{pseudogroup}.

\begin{figure}[ht]
  \centering
  \includegraphics[width=14cm]{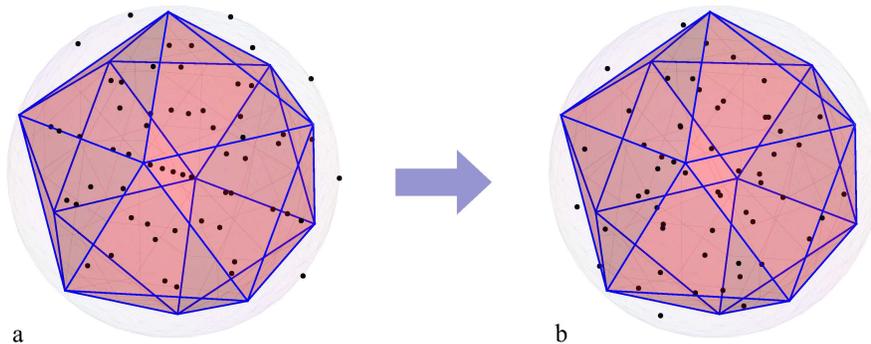}
  \caption{\textbf{a}: The icosahedral group representation inside the
    SU(2) sphere. \textbf{b}: the pseudogroup representation
    $\tilde{\mathcal{I}}(8)$ used in the preprocessor. Due to the
    large errors ($\sim 0.24$) the elements of
    $\tilde{\mathcal{I}}(8)$ seem to span the SU(2) sphere
    randomly.}
  \label{fig:pseudogr}
\end{figure}

Thanks to the homomorphism between SU(2) and $SO(3)$ we associate to
every rotation $ g \in \mathcal{I}$ a $2\times 2$ matrix
(\ref{oper}). Then we apply a brute-force search of length $L$ to
approximate the 60 elements in $\mathcal{I}$; in this way we obtain 60 braids that
give rise to the pseudogroup $\tilde{\mathcal{I}}(L) = \{
\tilde{g}_0(L),\tilde{g}_1(L), \dots, \tilde{g}_{59}(L) \}$. These braids
are characterized by an average distance $\epsilon(L)$ with their
corresponding elements $g_i \in \mathcal{I}$ given by
Eq.~(\ref{BFerror}). We notice from Fig. \ref{fig:pseudogr} that the
large errors for $L=8$ completely spoil the symmetry of the group (and
we will exploit this characteristic in the preprocessor of the hashing
algorithm); however, increasing the length $L$, one obtains
pseudogroups with smaller and smaller errors. Choosing, for instance,
a fixed braid length of $L = 24$, the error of each braid
representation to its corresponding exact matrix representation varies
from 0.003 to 0.094 with a mean distance of 0.018.

 
Therefore the braid representations of the icosahedral group with
different lengths are obtained by a brute-force search once and for
all. The so obtained braids are then stored for future utilizations
and this is the only step in which we apply, preliminarily, a
brute-force search.  Due to limiting computing resources, we construct
the pseudogroups representations for the 60 rotations up to the
length $L=68$, which, as we will describe below, is sufficient to
implement three iterations of the hashing algorithm. In principle one
could calculate the brute-force approximation of the group elements to
larger lengths once and for all, in order to use them for a greater
number of iterations. The average errors characterizing the main
pseudogroups we use in the iterations are shown in Fig. \ref{fig:bf};
they agree well with Eq.~(\ref{BFerror}).


Let us stress that that the 60 elements of $\tilde {\mathcal I}(L)$
(for any finite $L$) do not close any longer the composition laws of
the icosahedral group; in fact a pseudogroup $\tilde {\mathcal G}(L)$
becomes isomorphic to its corresponding group ${\mathcal G}$ only in
the limit $L \rightarrow \infty$.  If the composition law $g_i g_j =
g_k$ holds in $\mathcal{I}$, the product of the corresponding elements
$\tilde{g}_i(L)$ and $\tilde{g}_j(L)$ is not $\tilde{g}_k(L)$,
although it can be very close to it for large enough
$L$. Interestingly, the distance between the product $\tilde{g}_i(L)
\,\tilde{g}_j(L)$ and the corresponding element $g_k$ of ${\mathcal
  I}$ can be linked to the Wigner-Dyson distribution (see
Sec. \ref{WD}).

Using the pseudogroup structure of $\tilde {\mathcal
  I}(L)$, it is easy to generate a mesh $\mathcal{S}(L)$ made of a large
number of braids only in the vicinity of the identity matrix: this is
a simple consequence of the original group algebra, in
which the composition laws allow us to obtain the identity group
element in various ways. The set $\mathcal{S}$ is instrumental to
achieve an important goal, i.e. to search among the elements of $\mathcal{S}$
the best correction to apply to a previous approximation of the
target single-qubit gate $T$ we want to hash. 

\begin{figure}[th]
  \centering
  \includegraphics[width=14cm]{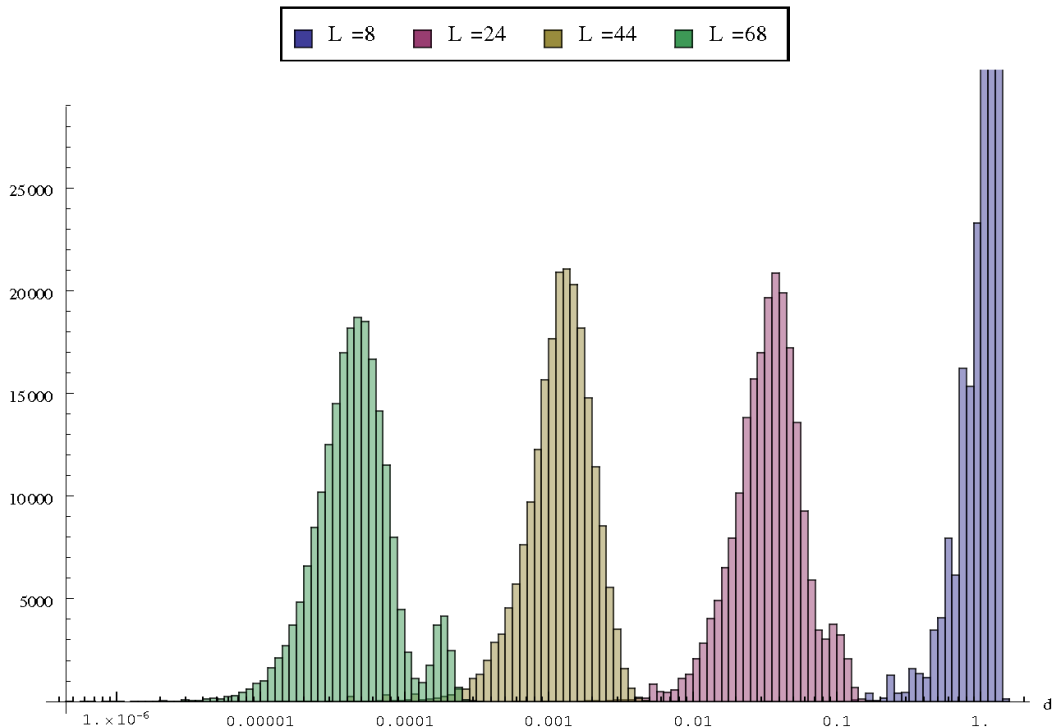}
  \caption{Distribution of the meshes $\mathcal{S}(L,3)$ used by the
    renormalization scheme of the hashing algorithm. The distance $d$
    from the identity is represented in logarithmic scale. The mesh
    with $L=8$ comprehends every possible product of three elements in
    $\tilde{\mathcal{I}}(8)$ and, therefore, it spans the whole
    SU(2) space up to the distance $\sqrt{2}$. The meshes
    $\mathcal{S}(24,3)$, $\mathcal{S}(44,3)$ and $\mathcal{S}(68,3)$,
    instead, are the product of four braids whose corresponding
    rotations are combined to approximate the
    identity. $\mathcal{S}(24,3)$ and $\mathcal{S}(44,3)$ follow the
    Wigner-Dyson distribution while $\mathcal{S}(68,3)$ exhibits a
    second local maximum due to the incomplete brute-force search we
    used to obtain $\tilde{\mathcal{I}}(68)$.}
  \label{fig:hist}
\end{figure}

It is important to notice that, changing the length $L$ of the
pseudogroup representation, we can control the average distance of
the fine rotations in $\mathcal{S}$ from the identity. To correct an
approximation of $T$ with an error $\varepsilon$, we need a mesh
$\mathcal{S}$ characterized by roughly the same average error in order
to reach an optimal density of possible corrections and so increase the efficiency
of the algorithm. Therefore, knowing the average error of the
distribution of the approximations we want to improve, we can choose a
suitable $L$ to generate a mesh. As represented in
Fig. \ref{fig:hist}, this allows us to define a series of denser and
denser meshes to iterate the hashing algorithm in order to correct at
each step the expected mean error, which we have determined by
analyzing the distributions of errors of 10000 random targets after
the corresponding iteration.

To create the mesh of fine rotations, labeled by $\mathcal{S}(L,n)$,
we consider all the possible ordered products $\tilde{g}_{i_1}(L)
\tilde{g}_{i_2}(L) \dots \tilde{g}_{i_n}(L)$ of a fixed $n\ge 2$
elements of $\tilde{\mathcal{I}}(L)$ of length $L$ and multiply them
by the matrix $\tilde{g}_{i_{n+1}}(L) \in \tilde{\mathcal{I}}(L)$ such
that $g_{i_{n+1}}=g_{i_n}^{-1}\dots g_{i_2}^{-1}g_{i_1}^{-1}$. In this
way we generate all the possible combinations of $n+1$ elements of
${\mathcal I}$ that produce the identity, but, thanks to the errors
that characterize the braid representation $\tilde{\mathcal{I}}$, we
obtain $60^n$ small rotations in SU(2), corresponding to braids of
length $(n+1)L$. In Sec. \ref{WD} we will describe their distribution
around the identity with the help of random matrices.

\section{Itarative pseudogroup hashing in SU(2)} \label{hashing}

\subsection{The iterative pseudogroup hashing algorithm}

The hashing algorithm is based on the possibility of finding
progressive corrections to minimize the error between the target gate
$T \in SU(2)$ and the braids that represent it. These corrections are
chosen among the meshes $\mathcal{S}(L_i,3)$ whose distribution around
the identity operator is shown in Fig. \ref{fig:hist}.

The algorithm consists of a first building block, called
\textit{preprocessor}, whose aim is to give an initial approximation
$\tilde T_0$ of $T$, and a \textit{main processor} composed of a
series of iterations of the hashing procedure that, at each step,
extend the previous representation by a braid in
$\mathcal{S}(L_i,3)$. The final braid has the form
\begin{equation} \label{out} \fl \tilde{T}_3 =
  \underbrace{\tilde{g}_{j_1}\left(L_0\right) \cdots
    \tilde{g}_{j_3}\left(L_0\right)}_{\rm
    Preprocessor}\underbrace{\tilde{g}_{p_1}\left(L_1\right) \cdots
    \tilde{g}_{p_4}\left(L_1\right)}_{\rm 1st\ Iteration}
  \underbrace{\tilde{g}_{q_1}\left(L_2\right) \cdots
    \tilde{g}_{q_4}\left(L_2\right)}_{\rm 2nd\ Iteration}
  \underbrace{\tilde{g}_{r_1}\left(L_3\right) \cdots
    \tilde{g}_{r_4}\left(L_3\right)}_{\rm 3rd\ Iteration}
\end{equation}
Each $\tilde{g}_{j}(L_i)$ is an element of the pseudogroup
$\tilde{\mathcal{I}}\left(L_i \right)$ and, as explained in the
previous section, the braid segment in each main iteration is
constrained by $g_{k_4}=g_{k_3}^{-1} g_{k_2}^{-1}g_{k_1}^{-1}$, $k =
p$, $q$, or $r$.

Each iteration starts from an input approximation $\tilde T_{i-1}$
with a distance $\varepsilon_{i-1}$ from the target $T$. We exploit
the elements of the mesh $\mathcal{S}(L_i,n)$ to generate a new braid
$\tilde T_{i}$ with a smaller distance $\varepsilon_{i}$. The lengths
$L_i$ in Eq.~(\ref{out}), which characterize the pseudogroups used in
the main processor, control the density of the corresponding meshes
and are chosen among the sets of stored pseudogroups in order to
correct the residual error in an efficient way (see
Sec. \ref{efficiency}).

Let us analyze now the details of each step in the hashing
algorithm. The preprocessor is a fast procedure to generate a rough
approximation of the target gate $T \in SU(2)$ and, in general, it
associates to every $T$ a braid which is an element of
$[\tilde{\mathcal{I}}\left(L_0\right)]^m$ (of length $m
L_0$). Therefore, the preprocessor approximates $T$ with the ordered
product of elements in the icosahedral pseudogroup
$\tilde{\mathcal{I}}\left(L_0\right)$ that best represents it, minimizing their distance. Thus we obtain
a starting braid
\begin{equation}
  \tilde T^{L_0,m}_0 = \tilde{g}_{j_1}\left(L_0\right)
  \tilde{g}_{j_2}\left(L_0\right) \dots
  \tilde{g}_{j_m}\left(L_0\right)
\end{equation}
with an initial error to reduce. The preprocessor procedure relies on
the fact that, choosing a small $L_0$, we obtain a substantial
discrepancy between the elements $g_i$ of the icosahedral group and
their representatives $\tilde g_i$, as shown in
Fig. \ref{fig:pseudogr}. Because of these seemingly random errors, the
set $[\tilde{\mathcal{I}}(L_0)]^m$ of all the products
$\tilde{g}_{j_1} \tilde{g}_{j_2} \dots \tilde{g}_{j_m}$ is well spread
all over SU(2) and can be thought as a random discretization of the
group. In particular we find that the pseudogroup
$\tilde{\mathcal{I}}(8)$ has an average error of about $0.24$ and it
is sufficient to take $m=3$ [as we did in Eq. (\ref{out})] to cover
the whole SU(2) in an efficient way with $60^3$ operators. The
average error for an arbitrary single-qubit gate with its nearest
element $\tilde T^{8,3}_0 \in [\tilde{\mathcal{I}}(8)]^3$ is about
0.027.

One can then apply the main processor to the first approximation
$\tilde T_0$ of the target gate. Each subsequent iteration 
improves the previous braid representation of $T$ by adding a finer
rotation to correct the discrepancy with the target and generate a new
braid.  In the first iteration we use the mesh $\mathcal{S}(L_1,n)$ to
efficiently reduce the error in $\tilde {T}^{l,m}_0$. Multiplying
$\tilde{T}^{l,m}_0$ by all the elements of $\mathcal{S}(L_1,n)$, we
generate $60^n$ ($O^n$ if we use a subgroup of order $O$) possible
braid representations of $T$ :
\begin{equation} \label{approx}  
\tilde{T}^{l,m}_0 \tilde{g}_{i_1} \tilde{g}_{i_2} \dots
  \tilde{g}_{i_n} \tilde{g}_{i_{n+1}} 
\end{equation}
Among these braids of length $(n+1)L_1 + mL_0$, we search the one with
the shortest distance to the target gate $T$. This braid,
$\tilde{T}_1\left(L_0,m,L_1,n\right) $, is the result of the first
iteration in the main processor.

The choice of $L=24$ for the first step is dictated by the analysis of
the mean error of the preprocessor ($\sim 0.03$) that requires, as we
will see in the following section, a pseudogroup with compatible error
for an efficient correction.  An example of the first iteration is
illustrated in Ref. \cite{burrello}.


With $\tilde{T}_1$ we can then apply the second and third iterations
of the main processor to obtain an output braid of the form in
Eq.~(\ref{out}). These iterations further reduce the residual
discrepancy in decreasing error scales. Each step in the main
processor requires the same runtime, during which a
search within $60^n$ braids selects the one with the shortest distance
to $T$. One must choose appropriate pseudogroups with longer braid
lengths $L_2$ and $L_3$ to generate finer meshes. As for $L_1$, we
choose $L_2$ and $L_3$ to match the error of the corresponding
pseudogroup to the respective mean residual error. In practice, we
choose $L_2 = 44$ and $L_3 = 68$. The final output assumes the form in
Eq.~(\ref{out}) and the average distance to the target braid (in 10000
random tests) is $2.29 \times 10^{-5}$ after the second iteration and
$8.24 \times 10^{-7}$ after the third. Without reduction the final
length is 568; however, due to shortenings at the junctions where
different braid segments meet, the practical final length of the weave
is usually smaller.

\begin{figure}
\begin{minipage}[c]{7.5cm}
\includegraphics[width=7.5cm]{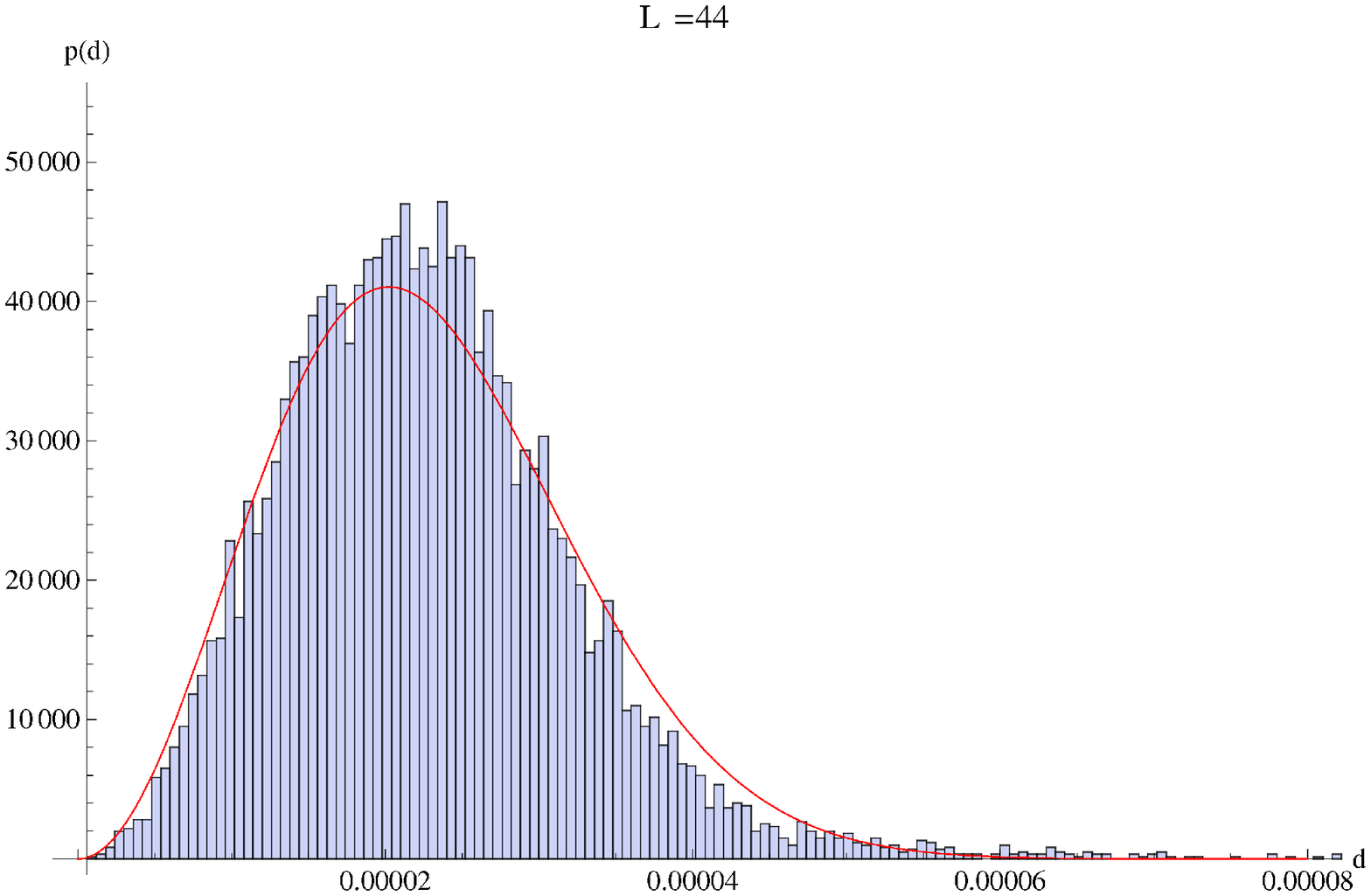}
\end{minipage}
\begin{minipage}[c]{7.5cm}
\includegraphics[width=7.5cm]{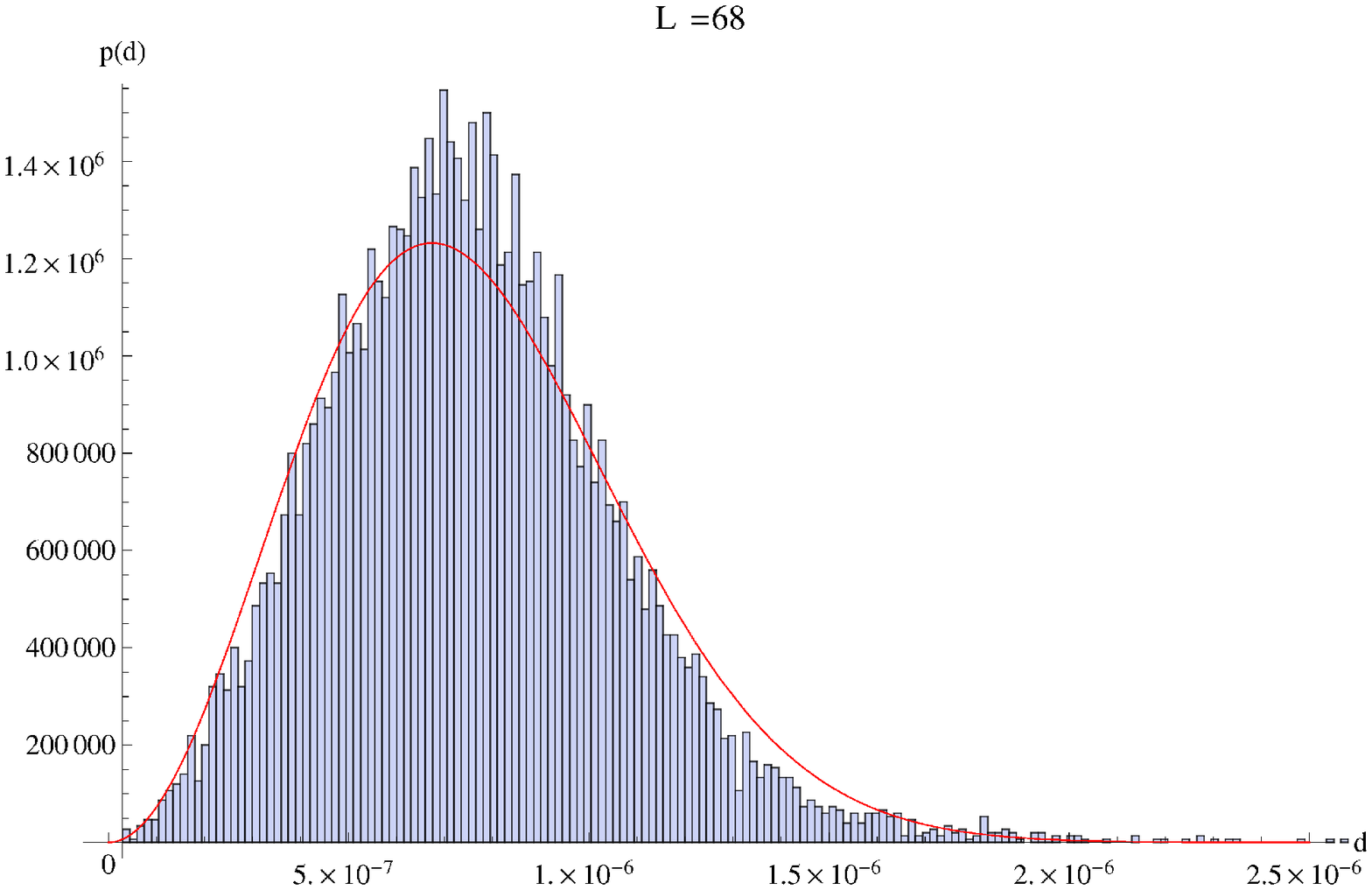}
\end{minipage}
\caption{Probability densities of the distance $d$ from the target of
  10000 random tests after the second ($L=44$) and third ($L=68$) main
  processors. The trend agrees with the unitary Wigner-Dyson
  distribution with average errors $2.28 \times 10^{-5}$ and $7.60
  \times 10^{-7}$, respectively. }
\label{fig:results}
\end{figure}

\subsection{Connection with random matrix theory and reduction factor 
for the main processor} 
\label{WD}

To analyze the efficiency of the main processor we must study the
random nature of the meshes $\mathcal{S}(L,n)$. The distribution of
the distance between the identity and their elements has an intriguing
connection to the Gaussian unitary ensemble of random matrices, which
helps us to understand how close we can approach the identity in this
way, and therefore what the optimal choice of the lengths of the
pseudogroups is for each iteration of the main processor.

Let us analyze the group property deviation for the 
pseudogroup $\tilde{\mathcal{I}}(L)$ for braids of length $L$. One
can write $\tilde{g}_i = g_i \me^{i \Delta_i}$, where $\Delta_i$ is a
Hermitian matrix, indicating the small deviation of the finite braid
representation to the corresponding SU(2) representation for an
individual element.  For a product of $\tilde{g}_i$ that approximate
$g_i g_j \cdots g_{n+1}=e$, one has
\begin{equation} \label{random}
\tilde{g}_i \tilde{g}_j  \cdots \tilde{g}_{n+1} = 
g_i \me^{i \Delta_i} g_j \me^{j \Delta_j}  
\cdots g_{n+1} \me^{i \Delta_{n+1}}
= \me^{i H_{n}},
\end{equation}
where $H_{n}$, related to the accumulated deviation, is 
\begin{equation}
\label{eq:rg}
\fl H_{n} = g_i \Delta_i g_i^{-1} + g_i g_j \Delta_j g_j^{-1} g_i^{-1}
+ \cdots + g_i g_j \cdots g_n\Delta_n g_n^{-1} \cdots g_j^{-1}g_i^{-1}
+\Delta_{n+1}+ O(\Delta^2).
\end{equation}
The natural conjecture is that, for a long enough sequence of matrix
product, the Hermitian matrix $H_n$ tends to a random matrix
corresponding to the Gaussian unitary ensemble. This is plausible as
$H_n$ is a Hermitian matrix that is the sum of random initial
deviation matrices with random unitary transformations. A direct
consequence is that the distribution of the eigenvalue spacing $s$
obeys the Wigner-Dyson form~\cite{mehta},
\begin{equation} \label{Wigner}
P(s) = {\frac{32}{\pi^2 s_0}} \left ({\frac{s}{s_0}}\right )^2 
\me^{-(4/\pi) (s/s_0)^2},
\end{equation}
where $s_0$ is the mean level spacing.  For small enough deviations,
the distance of $H_n$ to the identity, $d \left (1, \me^{iH_n} \right
) = \Vert H_n \Vert + O \left (\Vert H_n \Vert^3 \right )$, is
proportional to the eigenvalue spacing of $H$ and, therefore,
should obey the same Wigner-Dyson distribution. The conjecture above
is indeed well supported by our numerical analysis, even for $n$ as
small as 3 or 4: the distances characterizing the meshes with $L=24$ and $L=48$ obtained from the corresponding pseudogroups (Fig. \ref{fig:hist}) follow the unitary Wigner-Dyson distribution. 


The elements of the meshes $\mathcal{S}(L,n)$ are of the form in
Eq.~(\ref{random}) and this implies that, once we choose a pseudogroup
$\tilde{\mathcal{I}}(L)$ whose average error $\bar d(L)$ is given by
Eq.~(\ref{BFerror}), the mean distance $s_0$ in (\ref{Wigner}) of the
corresponding mesh from the identity is $s_0(L) \approx
\sqrt{n+1}\,\bar{d}(L)$ as resulting from the sum of $n+1$ gaussian
terms.



At each iteration of the main processor, we increase the braid by $(n
+ 1) = 4$ braid segments with length $L_i$.  By doing that, we create
$60^n$ braids and the main processor search, among them, the best
approximation of the target. Therefore, the runtime is linear in the
dimension of the meshes used and in the number of iterations. The
unitary random matrix distribution implies that the mean deviation of
the 4-segment braids from the identity (or any other in its vicinity)
is a factor of $\sqrt{n+1}$ times larger than that of an individual
segment. Considering the 3-dimensional nature of the unitary matrix
space, we find that at each iteration the error (of the final braid to the target gate)
is reduced, on average, by a factor of $f \sim 60^{n/3}/\sqrt{n+1} =
30$, where 60 is the number of elements in the icosahedral group. This
has been confirmed in the numerical implementation. 

\subsection{Hashing with the cubic group}

For comparison, we also implemented the hashing procedure with the
smaller cubic group. In this case the rotations in the group of the
cube are 24; thus we choose $n=4$ to generate a comparable number of
elements in each mesh $\mathcal{S}(L,n)$. Our implementation of the
hashing with the cubic group uses a preprocessor with $L_0=8$
and $m=4$ and a main processor with $L_1=24$ and $n=4$. Approximating
over 100 random targets, we obtained an average error $6.92 \times
10^{-4}$ after the main processor, comparable to $7.24 \times 10^{-4}$
after the first iteration in the previous icosahedral group
implementation. This result is consistent with the new reduction
factor $f_{\rm cube}=24^{4/3}/\sqrt{5}\approx 30.96$. However we note
that the cubic hashing is less efficient both in terms of the braid
length (because it requires $n=4$ instead of $n=3$) and in terms of
the runtime (because the time required for the searching algorithm is
linear in the elements of $\mathcal{S}$ and $24^4 > 60^3$).

\subsection{Tail correction}
\label{sec:tail}

The choice of the proper $L_i$ is important. We determine them by the
average error before each iteration. If a certain $L_i$ is too large,
it generates a mesh around the identity that may be too small to
correct the error relatively far from the identity, where the mesh is
very fine. On the other hand, if $L_i$ is too small, the mesh may be
too sparse to correct efficiently. The former situation occurs exactly
when we treat the braids with errors significantly larger than the
average; they correspond to the rare events in the tail of the
distributions as shown in Fig.~\ref{fig:results}. Such an already
large error is then amplified with the fixed selection of $L_i=24$, 44,
and 68.  To avoid this, one can correct the ``rare'' braids
$\tilde{T}_{i-1}$, whose error is higher than a certain threshold,
with a broader mesh [e.g., $\mathcal{S}(L_i-4,n)$].

The tail correction is very efficient. If we correct for the $0.6\%$
of the targets with the largest errors in the second iteration with
$\mathcal{S}(40,3)$ instead of $\mathcal{S}(44,3)$, we reduce the
average error by about $8\%$ after the third iteration (see Table
\ref{tab:results}). The drastic improvement is due to the fact that
once a braid is not properly corrected in the second iteration, the
third one becomes ineffective. We can illustrate this situation with
the example of the operator $iY$ (where $Y$ is the Pauli matrix):
without tail correction it is approximated in the first iteration
with an error of $0.0039$, which is more than 5 times the average error
expected. After the second iteration, we obtain an error of
$4.3 \times 10^{-4}$ (almost 20 times higher than the average value)
and after the third the error becomes $2.0 \times 10^{-4}$ (more
than 200 times the mean value). If we use $\mathcal{S}(40,3)$ instead
of $\mathcal{S}(44,3)$ in the second iteration we obtain an error
$4.46 \times 10^{-5}$, with a shorter braid than before, and a final
error of $1.31 \times 10^{-6}$ which is less than twice the average
error.

\begin{table}
\centering
\begin{tabular}{|c|c|c|c|c|}
\hline
& \multicolumn{2}{|c|}{Without tail correction} & \multicolumn{2}{|c|}{With tail correction} \\
\hline
10000 trials & Average Error & $\sigma$ & Average Error & $\sigma$ \\ \hline
Preprocessor* & $0.027$ & $0.010$ & $0.027$ & $0.010$ \\
Main, 1st iteration* & $7.24 \times10^{-4}$ & $3.36 \times10^{-4}$ & $7.24 \times10^{-4}$ & $3.36 \times10^{-4}$\\
Main, 2nd iteration &  $2.29 \times10^{-5}$ & $1.3 \times10^{-5}$ & $2.28 \times10^{-5}$ & $9.79 \times10^{-6}$\\
Main, 3rd iteration & $8.24 \times10^{-7}$ & $5.6 \times10^{-6}$ & $7.60 \times10^{-7}$ & $3.27 \times10^{-7}$\\
\hline
\end{tabular}
\caption{Average error and standard deviation for the hashing
  algorithm after the preprocessor and the three iterations of the
  main processor. The outputs in absence or presence of a tail
  correction for the second and third iterations are shown (the
  asterisk indicates that the preprocessor and the first iteration are
  not affected by the tail correction). This correction is based on
  the psudogroups with length 40 and 64 instead of the standard ones,
  44 and 68. Only about the $0.6\%$ of the targets used the tail
  correction in the second iteration, but one notice that the results,
  both in terms of average error and in terms of standard deviation
  $\sigma$, are extremely affected by these rare events.}
\label{tab:results}
\end{table}

\section{General efficiency of the algorithm} \label{efficiency}

To evaluate the efficiency of the hashing algorithm it is useful to
calculate the behaviour of the maximum length of the braids and of the
runtime with respect to the average error obtained. We compare the
results with those of the brute-force search (which gives the optimal
braid length) and of the Solovay-Kitaev algorithm.

As described in Sec.~\ref{WD}, we reduce the average error at the
$i$th iteration to
\begin{equation} \label{eqa}
\varepsilon_i \sim \varepsilon_{i-1} /f \qquad {\rm with } \quad f\approx 30.
\end{equation}
The total number of iterations (or {\it depth}) to achieve a final
error of $\varepsilon$ is then
\begin{equation}
  M \sim \frac{\ln (1/\varepsilon)}{\ln{f}},
\end{equation}
so the expected error after each iteration is
\begin{equation}
\ln (1/\varepsilon_i) \sim i \ln f. 
\end{equation}

\begin{figure}
 \centering
 \includegraphics[width=12cm]{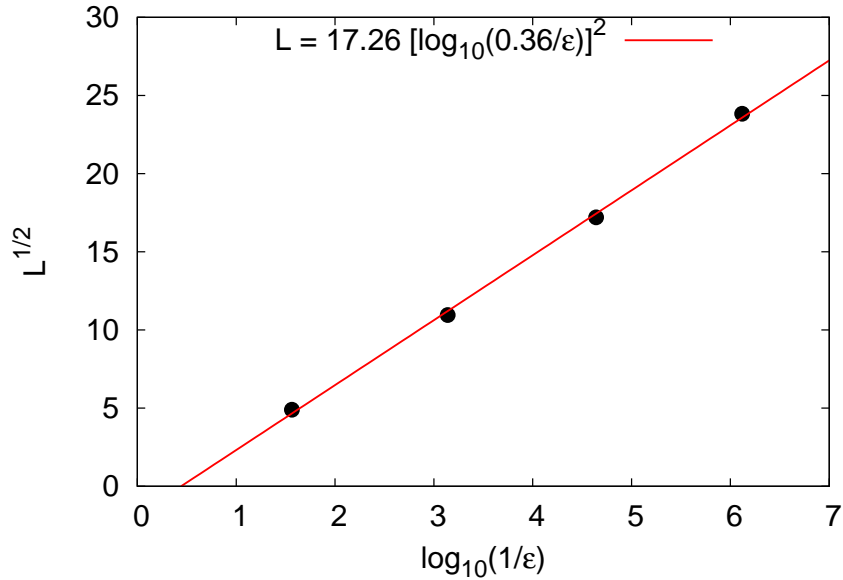}
 \caption{The scaling performance of the hashing algorithm in terms of
   the square root of the maximum length versus the logarithm of the
   inverse error. The error is the average error in approximating
   10000 random targets after the preprocessor and the three
   iterations in the main processor. The results agree with the
   expected behaviour [Eqs.~(\ref{eq1}) and (\ref{eq2})].}
\label{fig:performance}
\end{figure}

For efficient optimization, we choose the length $L_i$ of the braid
segments at the $i$th iteration to approximate the corresponding
icosahedral group elements with an average error of
$\varepsilon_{i-1}$ (see discussions in Sec.~\ref{sec:tail}). So we
have, from Eq.~(\ref{BFerror}),
\begin{equation}
L_i \sim \mathcal{L} \ln \left( {1/\varepsilon_{i-1}}\right) 
\end{equation}
with $\mathcal{L}\approx 6$ [see Eq.~(\ref{BFerror})] and the length
of the braid we construct increases by $(n+1)L_i = 4 L_i$ at the
$i$-th iteration, i.e.,
\begin{equation}
L_i - L_{i-1} = 4 \mathcal{L} \ln\left( {1/\varepsilon_{i-1}}\right) 
\sim 4 \mathcal{L} (i - 1) \ln f.
\end{equation}
Thus the total length of the braid with an error of $\varepsilon$ is
\begin{equation}
L_M \sim \sum_{i=1}^M 4 \mathcal{L} (i - 1) \ln{f} \sim M^2.
\end{equation}
The final results for the hashing algorithm are, therefore,
\begin{eqnarray}
L_{\rm qh} \sim \left( \ln\left(1/\varepsilon \right) \right)^2,
\label{eq1} \\
T_{\rm qh} \sim M \sim \ln\left(1/\varepsilon \right). \label{eq2}
\end{eqnarray}
We have explicitly confirmed that we can realize the perfect
length-error scaling as shown in Fig.~\ref{fig:performance} up to
three iterations in the main processor.

We can conclude that while no method beats the brute-force search in
length, we achieve a respectable gain in time. Comparing these results
with the length of the braids obtained by the Solovay-Kitaev algorithm
in Eq.~(\ref{LKS}) and with its runtime in Eq.~(\ref{TKS}), the
hashing algorithm gives results that are better than the
Solovay-Kitaev results in terms of length and significantly better in
terms of time.

\section{Conclusions}
\label{sec:conclusion}

We have demonstrated, for a generic universal topological quantum
computer, that the iterative pseudogroup hashing algorithm allows an
efficient search for a braid sequence that approximates an arbitrary
given single-qubit gate. This can be generalized to the search for
two-qubit gates as well. The algorithm applies to the quantum gate
construction in a conventional quantum computer given a small
universal gate set, or any other problems that involve realizing an
arbitrary unitary rotation approximately by a sequence of
``elementary'' rotations.

The algorithm uses a set of pseudogroups based on the icosahedral
group or other finite subgroups of SU(2), whose multiplication
tables help generate, in a controllable fashion, smaller and smaller
unitary rotations, which gradually (exponentially) reduce the distance
between the result and the target. The iteration is in the same spirit
as a generic renormalization group approach, which guarantees that the
runtime of the algorithm is proportional to the logarithm of the
inverse average final error $1 / \varepsilon$; the total length of the
braid is instead quadratic in $\ln (1 / \varepsilon)$, and both the
results are better than the Solovay-Kitaev algorithm introduced in
textbooks.

We have explicitly demonstrated that the result from the performance
analysis is in excellent agreement with that from our computer
simulation. We also showed that the residual error distributes
according to the Wigner-Dyson unitary ensemble of random matrices. The
connection of the error distribution to random matrix theory ensures
that we can efficiently carry out the algorithm and improve the rare
cases in the distribution tail.

The overhead of the algorithm is that we need to prepare several sets
of braid representations of the finite subgroup elements. Obtaining
the longer representation can be time-consuming; but fortunately, we
only need to compute them once and use the same sets of representations for
all future searches.

\section{Acknowledgements}

This work is supported by INSTANS (from ESF), 2007JHLPEZ (from MIUR),
and the 973 Program under Project No. 2009CB929100.
X. W. acknowledges the Max Planck Society and the Korea Ministry of
Education, Science and Technology for the support of the Independent
Junior Research Group at the APCTP. M. B. and G. M. are grateful to
the APCTP for the hospitality in Pohang where part of this study was
carried out.

\appendix

\section{Distribution of the best approximation in a given set of
  braids}
\label{sec:weave}

In this Appendix we derive the distribution of the approximation to a
gate in a given set of $N$ braids in the vicinity of the
identity. While it is sufficient for the discussions in the main text,
this derivation can be generalized to the more generic cases. Let us
assume that the targeted gate is $g_0 = U(\hat{l},\phi_0)$ as defined
in Eq.~(\ref{oper}).  The distance between $g_0$ and the identity is
$s_0 = 2 \sin(\phi_0/4) \approx \phi_0/2$ for small $\phi_0$. We then
search in a given set of braids, either with a distribution as given
in Eq.~(\ref{pBF}) or from a generated random matrix distribution as
discussed in Sec. \ref{WD}, the one with the shortest distance to the
target.

We consider an arbitrary braid with representation $g = U\left(\hat{m},
\phi\right)$ in a collection with a distribution $p(s)$ of the
distance to the identity $s = 2 \sin(\phi/4)$. We define
\begin{equation}
P(x) = \int_0^x p(s) ds,
\end{equation}
which is the probability of having a distance less than $x \le
\sqrt{2}$.  Obviously $P(x)$ is a monotonically increasing function
bound by $P(0) = 0$ and $P(\sqrt{2}) = 1$.  The distance $d(g_0, g)$
between $g$ and $g_0$ is the same as that between $g_0^{-1} g$ and the
identity. To the first order in $\phi$ and $\phi_0$ (as we assume all
braids/gates are in the vicinity of the identity braid) we have, from
Eq.~(\ref{dist})
\begin{equation}
d(g_0, g) \approx \sqrt{s^2 + s_0^2 - 2 (\hat{l} \cdot \hat{m}) s_0 s}
= \left \vert \left \vert s \hat{m} - s_0 \hat{l}
\right \vert \right \vert,
\end{equation}
which is bound by $s+s_0$ and $\vert s - s_0 \vert$.
We can see that the chance to find a braid that is close to
$g_0$ is large when $p(s)$ peaks around $s_0$. If we denote the
probability of having no braids within a distance of $t$ by $Q(t)$,
the probability of having the braid with the shortest distance between
$t$ and $t + dt$ is
\begin{equation}
Q(t) - Q(t+dt) = Q(t) 
\left \langle N dt 
\int_0^{\sqrt{2}} p(s) ds \delta \left (t - d(g_0, g) \right )
\right \rangle ,
\end{equation}
where the angled bracket implying the angular average of the
averaged number of braids with a distance between $t$ and $t+dt$.  
Therefore, 
\begin{equation}
- \frac{d \ln Q(t)}{dt} = N
\left \langle 
\int_0^{\sqrt{2}} p(s) ds \delta \left (t - d(g_0, g) \right )
\right \rangle .
\end{equation}
As an example, we consider $s_0 = 0$ [i.e., with the full SU(2)
  rotation symmetry of the distribution], in which case
\begin{equation}
- \ln Q(t) = N \int_0^t p(s) ds = N P(t),
\end{equation}
or $Q(t) = \exp[-N P(t)]$. $NP(t)$ is the expected number of braids
with a distance to the identity less than $t$. The differential
probability of having the braid with the nearest distance between $t$
and $t+dt$ is, therefore,
\begin{equation}
q(t) \equiv - {\frac{dQ(t)}{dt}} = N {dP(t) \over dt} e^{-N P(t)}
= N p(t) e^{-N P(t)}.
\end{equation}
Combining the results with Eq.~(\ref{pBF}), we estimate for
the brute-force search
\begin{equation}
q_{BF}(t; L) = N(L) p_{BF}(t) e^{-N(L) P_{BF}(t)}
\end{equation}
with an average distance
\begin{equation}
\bar{d}(L) = \frac{\pi^{1/3} \left[\Gamma
   \left(\frac{1}{3}\right)-\Gamma \left(\frac{1}{3},\frac{8
   \sqrt{2} N(L)}{3 \pi }\right)\right]}{6^{2/3} [N(L)]^{1/3}}
\end{equation}
where $\Gamma(a,x)$ is the incomplete gamma function 
\begin{equation}
\Gamma(a,x) = \int_x^{\infty} dt t^{a-1} e^{-t}.
\end{equation} 
In the large $L$ limit, $\bar{d}(L) \approx 1.021 e^{-L/5.970}$. This
is the result we quoted in Eq.~(\ref{BFerror}).

\end{document}